\documentclass[iop]{emulateapj}
\usepackage{txfonts}
\usepackage{graphicx}
\usepackage{natbib}

\newcommand{\farcss}{\mbox{\ensuremath{^{\prime\prime}}}}

\def\etal{{\sl et al.}}
\usepackage{natbib}
\shorttitle{Magnetic Flux Cancellation in Ellerman Bombs}
\shortauthors{Reid et al.}

\begin{document}

\title{Magnetic Flux Cancellation in Ellerman Bombs}
\vskip1.0truecm
\author{
A. Reid$^{1,2}$, M. Mathioudakis$^{1}$, J. G. Doyle${^2}$, E. Scullion$^{3}$, C. J. Nelson$^{1,4}$, V. Henriques$^1$, T. Ray$^5$}
\affil{
1. Astrophysics Research Centre, School of Mathematics and Physics, Queen's University Belfast, BT7~1NN, Northern Ireland, UK; e-mail: areid29@qub.ac.uk\\
2. Armagh Observatory, College Hill, Armagh, BT61 9DG, UK\\
3. Trinity College Dublin, College Green, Dublin 2, Ireland\\
4. Solar Physics and Space Plasma Research Centre, University of Sheffield, Hicks Building, Hounsfield Road, Sheffield, UK, S3 7RH\\
5. Dublin Institute for Advanced Studies, 31 Fitzwilliam Place, Dublin 2, Ireland\\
}
%

\begin{abstract}
Ellerman Bombs (EBs) are often found co-spatial with bipolar photospheric magnetic fields. We use H$\alpha$ imaging spectroscopy along with Fe I 6302.5~\AA\ spectro-polarimetry from the Swedish 1-m Solar Telescope (SST), combined with data from the Solar Dynamic Observatory (SDO) to study EBs and the evolution of the local magnetic fields at EB locations. The EBs are found via an EB detection and tracking algorithm. We find, using NICOLE inversions of the spectro-polarimetric data, that on average (3.43 $\pm$ 0.49) x 10$^{24}$ ergs of stored magnetic energy disappears from the bipolar region during the EBs burning. The inversions also show flux cancellation rates of 10$^{14}$ - 10$^{15}$ Mx s$^{-1}$, and temperature enhancements of 200~K at the detection footpoints.  We investigate near-simultaneous flaring of EBs due to co-temporal flux emergence from a sunspot, which shows a decrease in transverse velocity when interacting with an existing, stationary area of opposite polarity magnetic flux and the EBs are formed. We also show that these EBs can get fueled further by additional, faster moving, negative magnetic flux regions.

\end{abstract}


\section{INTRODUCTION}

Ellerman Bombs (EBs) are small-scale, short-lived, impulsive brightenings originally detected in the outer wings of the H$\alpha$ line \citep{ell}. They occur exclusively near solar active regions \citep{zac, geo} or areas of enhanced photospheric magnetic activity (e.g \citep{par, wat}). The H$\alpha$ line profile of an EB has an absorption core, that remains unchanged relative to the local background, and emissions in the line wings during the lifetime of the EB. The emission wings may be asymmetric, which can be due to overlying chromospheric flows \citep{bruzek, kitai, dara, wat1, viss}. EBs also appear in Ca II 8542~\AA\ with line profile characteristics similar to H$\alpha$ \citep{fang, socas4, par2,viss,li}. EBs are thought to produce no observable effects in the upper atmosphere \citep{viss}, though there may be some tentative indications of heated areas of transition region above the location of EB activity \citep{schmieder}. Brightenings from EBs are also observable in the SDO 1700~\AA\ and 1600~\AA\ channels, though to a lesser degree than the H$\alpha$ line wings due to broad passbands of these filters encompassing a wide range of atmospheric heights \citep{viss}. While the 1600~\AA\ channel offers better contrast than the 1700~\AA\ channel \citep{rutt, viss}, EB signatures are more difficult to observe in the 1600~\AA\ channel due to contamination effects from C IV emission with transition region temperatures.

More recent studies involving the Interface Region Imaging Spectrograph explorer find `bombs' in atmospheric lines such as Si IV, C II and Mg II, indicating that these regions are host to pockets of hot plasma, with possible temperatures ranging from 6000-80,000K and bi-directional flows of up to 80 km s$^{-1}$ \citep{Peter}. These `bombs' have been observed co-spatially with EBs found in H$\alpha$ by \cite{viss2}, and now hint that the tops of EBs may be heated to transition region temperatures at physical heights below the chromospheric canopy, putting the previous temperature estimates from modeling EBs into question. \cite{judge} however has debated the origins of these `bombs', speculating that their formation is due to Alfv\'{e}nic turbulence in the low-mid chromosphere.

Several studies of EBs connect their detection in H$\alpha$ with regions of opposite polarity photospheric magnetic fields \citep{geo, wat, matsu, hashi, Nelson1, viss}. More recent studies also hint at possible flux cancellation of the bipoles at EB sites \citep{matsu, Nelson1, Nelson2} with values for flux cancellation in the region of 3 - 8.5 x $10^{14} $ Mx s$^{-1}$. It is thought that this flux cancellation in the form of photospheric magnetic reconnection is the driver for the appearance of EBs \citep{geo, iso, wat, matsu2, matsu, hashi}.  It has been shown numerically that photospheric magnetic reconnection would be most efficient at the temperature minimum at a height of 600~km above the lower photospheric boundary \citep{lit2}. EBs have been estimated to form at this height \citep{Nelson3}, with footpoints reported to form as low as the intergranular lanes, near the photospheric floor \citep{wat1}.

There are 3 main mechanisms related to EB events and their associated magnetic topologies. Two of these mechanisms involve reconnection between areas of opposite polarity magnetic flux. The first of these is triggered by the emergence of new flux interacting with an existing opposite polarity area \citep{wat, hashi}. The second mechanism involves reconnection along a resistive, undulatory ``sea serpent" flux emergence \citep{geo, par}. The final mechanism does not involve opposite polarity reconnection, and instead the EB is caused by shearing reconnection in a unipolar region of magnetic flux \citep{geo, wat}.

Three-dimensional numerical modeling of the ``sea serpent" reconnection case has been studied, showing a local temperature increase ratio in the photosphere of 1.1-1.5 relative to quiet Sun, along with a density increase by a factor of 4 at the reconnection site \citep{ach}. The \cite{ach} model has also shown bi-directional flows in the region, with values of 2-4 km s$^{-1}$.  Semi-empirical models for EBs show localized temperature enhancements of 600-3000K around the temperature minimum region \citep{fang, berlicki}. These temperature enhancements lead to intensity enhancements in the wings of the H$\alpha$ and Ca II 8542~\AA\ lines, while the line cores are formed higher in the chromosphere.  Other studies of EBs also find similar temperature enhancements ranging from 200-3000K in the photosphere/temperature minimum region \citep{geo, iso, yang,hong, li}.

Radiative energies of EBs have also been considered, by estimating the radiative loss rate from H$\alpha$. Assuming an EB lifetime of 600 seconds, with a depth of 100~km, and measuring the apparent area of the brightenings, \cite{geo} found that EBs have a total radiative energy of 10$^{27}$ - 10$^{28}$ ergs, with peak energy rates of 10$^{25}$ ergs s$^{-1}$. The statistical study of \cite{Nelson} applied a similar method to the results of an automated detection algorithm for EBs and found a lower total radiative energy of 10$^{22}$ - 10$^{25}$ ergs, with peak radiative loss rates of 10$^{21}$ - 10$^{23}$ ergs s$^{-1}$.

In this paper, we use high spatial and temporal resolution H$\alpha$ imaging spectroscopy along with Fe I 6302.5~\AA\ imaging spectro-polarimetry from the Swedish 1-m Solar Telescope to study EBs and their associated flux cancellation rates. The EBs are detected and tracked using an automated algorithm. The identified features are then inverted using NICOLE to produce estimated flux cancellation rates and temperature information.

\section{OBSERVATIONS AND DATA REDUCTION}

\begin{figure*}[!t]
\plotone{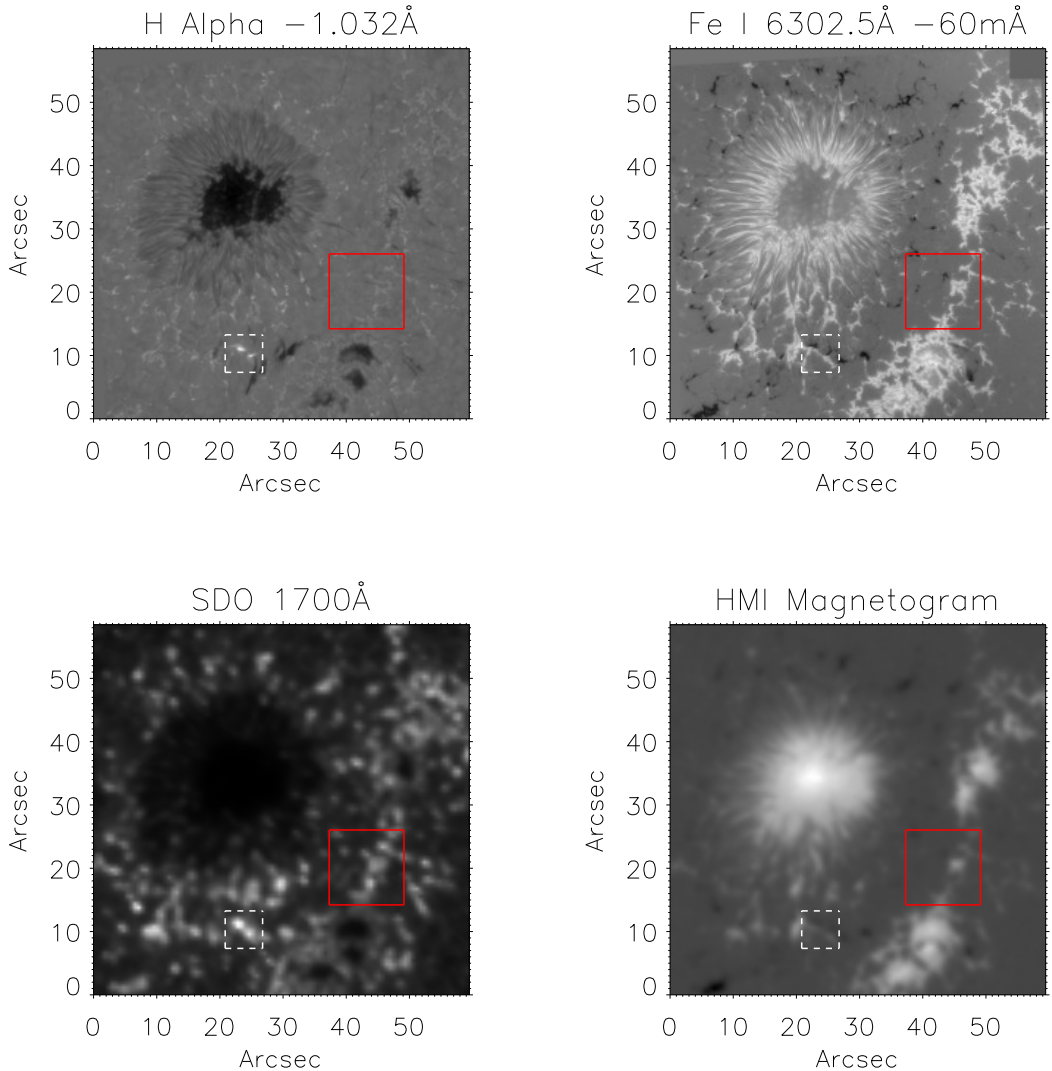}

\caption{Top Left: H$\alpha$ image -1.032~\AA\ from line centre showing the FOV of the SST.  Two flaring EBs can be seen (highlighted white dashed box). The red solid box shows the area of quiet-Sun used for reference. Top Right: Co-spatial and co-temporal Stokes-$V$ image. Bottom Left: Co-spatial SDO 1700~\AA\ snapshot. Bottom Right: Co-spatial HMI line of sight magnetogram.}
\label{Figure1}
\end{figure*}

The observations were carried out with the CRisp Imaging SpectroPolarimeter (CRISP) at the Swedish 1-m Solar Telescope \citep{sst, sst2} on La Palma.  
 The target was active region NOAA 12077, near disk centre (coordinates: X= 180\farcss, Y= -81\farcss, $\mu = 0.97$). The observations took place on 2014 June 5 between 08:27-09:58 UT. The observational setup comprised of H$\alpha$ line scans using 5 points of $\pm$ 1.032~\AA\ , $\pm$ 0.774~\AA\ and line core, imaging  spectro-polarimetry in Fe I 6302~\AA, sampled across 11 points, ranging from $\pm$ 0.15~\AA\ from line centre, in steps of 30m\AA. A Fe I scan was taken after every 9 H$\alpha$ scans. The spectro-polarimetric data had a post-reduction mean cadence of 45 seconds, while the H$\alpha$ spectral imaging had a mean cadence of 3.2 seconds, with 17 second cadence when the Fe I data was being acquired. The image scale of the observations was 0.059 \farcss per pixel, with a total field-of-view (FOV) of 59 \farcss x 58 \farcss. A snapshot of the FOV is shown in Fig.~\ref{Figure1}. 


The data were processed with the Multi-Object Multi-Frame Blind Deconvolution (MOMFBD) algorithm \citep{noort}. This includes tessellation of the images into 64x64 pixels$^2$ sub-images for individual restoration, done over each temporal frame and line position within the scans. Wide-band images act as a stabilizer for the narrow-band alignment causing the different polarimetric states to be consistent, seeing and reconstruction-wise, thus  preventing seeing-induced cross-talk during demodulation \citep{vasco}. Prefilter field-of-view and wavelength dependent corrections were applied to the restored images. The spectro-polarimetric data were also demodulated to remove the cross-talk between the Stokes parameters \citep{schnerr}.
The final correction involved the long-scale cavity error of the instrument.  Further information on MOMFBD image restoration techniques is available in \cite{noort2} and \cite{noort3}.\\

The SST observations were combined with data from the 1700~\AA\ passband of the Atmospheric Imaging Assembly \citep{lemen} and Helioseismic Magnetic Imager \citep{scherrer} on the Solar Dynamics Observatory (SDO). These were reduced, and cropped temporally to match the timestamps of the H$\alpha$ SST data cube as closely as possible for AIA. The HMI magnetograms were temporally aligned to the timestamps of the spectro-polarimetric data. The SDO data were then spatially aligned to the SST data. This was done by centering the SDO data cubes on the centre pointing value from the SST. The SDO data were scaled to match the pixel resolution of SST, and finally rotated with respect to the observation angle of the SST data. The accuracy of the alignment for the data was checked by comparing the central positions of the dark sunspot regions with the H$\alpha$ -1.032~\AA\ images. After this, a co-aligned datacube was made for the 1700~\AA\ AIA data and magnetograms, with finer calibrations to the alignment made manually for each channel, for the whole timeseries. Aligned frames of the full FOV are shown in Fig.~\ref{Figure1}.


\section{Ellerman Bomb Detection and Tracking}

The detection of EBs was carried out with an Ellerman Bomb Detection Automation and Tracking Algorithm (EBDATA). For a feature to be classified as an EB it has to fulfil the following criteria.  

\begin{enumerate}
\item{ The feature must have at least one pixel with intensity \textgreater 145\% that of the local quiet Sun in both wings of H$\alpha$ at $\pm$ 1.032~\AA\.. }
\item{ The surrounding area is grown to an intensity threshold of \textgreater 130\% using the same quiet Sun profile above at the same wavelength positions. The grown area has to be greater than 15 pixels.}
\item{ The line core in H$\alpha$ must remain unchanged (no more than 10\% increase to account for variability), relative to the average line core intensity at the EB location over the previous 60 seconds.}
\item{ The temporal variation of the intensity must show evidence of impulsivity (10\% increase in the intensity in the grown EB area over the previous 60 seconds).}
\item{ The lifetime of the event needs to be \textgreater 45 seconds.}
 \end{enumerate}

The first 2 criteria use intensity thresholding to identify possible candidates similar to \cite{viss}, though with lower intensities. The lower intensities were chosen because the average value is taken over an area of quiet Sun, not the full field-of-view, and so in discounting the sunspot, the relative average increases. The quiet Sun profile was taken over an area of 200 pixels$^{2}$ (11.84 \farcss $^{2}$), centered about the position X= 43.216\farcss, Y=20.128\farcss, seen in the red box of Fig.~\ref{Figure1}. The intensity threshold for the grown area is the same as that of \cite{Nelson}. The size criterion was added to ensure no small scale anomalies were picked up as detections. Detection criterion 4 calls for the potential EB to show impulsivity to ensure no moving magnetic features were falsely identified (pseudo-EBs; \cite{rutt}). EBs are impulsive reconnection events and should exhibit some form of flaring as one of their main signatures \citep{wat1, viss, rutt, reid}. This intensity change was to be only in the wings and not in the core of H$\alpha$. The intensity change is averaged over the whole grown area and is a running average.

If detection criteria 1- 3 are met, the detected area is placed into a detection binary cube. This datacube then runs through the tracking algorithm. The algorithm looks at the detections in each timeframe. The first frame containing detections will label each detection area in the binary map with a tracking number. Tracking numbers are only assigned if the impulsivity criterion is fulfilled. Subsequent frames are then scanned for individual detections, and the pixels within each detection are checked for any overlap with labelled EBs within the previous 60 seconds, to allow for lapses in seeing. If a detection shows any overlap with a previously labelled EB, then the detection is labelled with the overlapping tracking number (if multiple overlapped numbers, it takes the tracking number of the one with the highest correlation).  After this, a clean up routine is performed, which removes any detection with a lifetime less than 45 seconds (criterion 5).

\begin{figure*}[!t]
\plotone{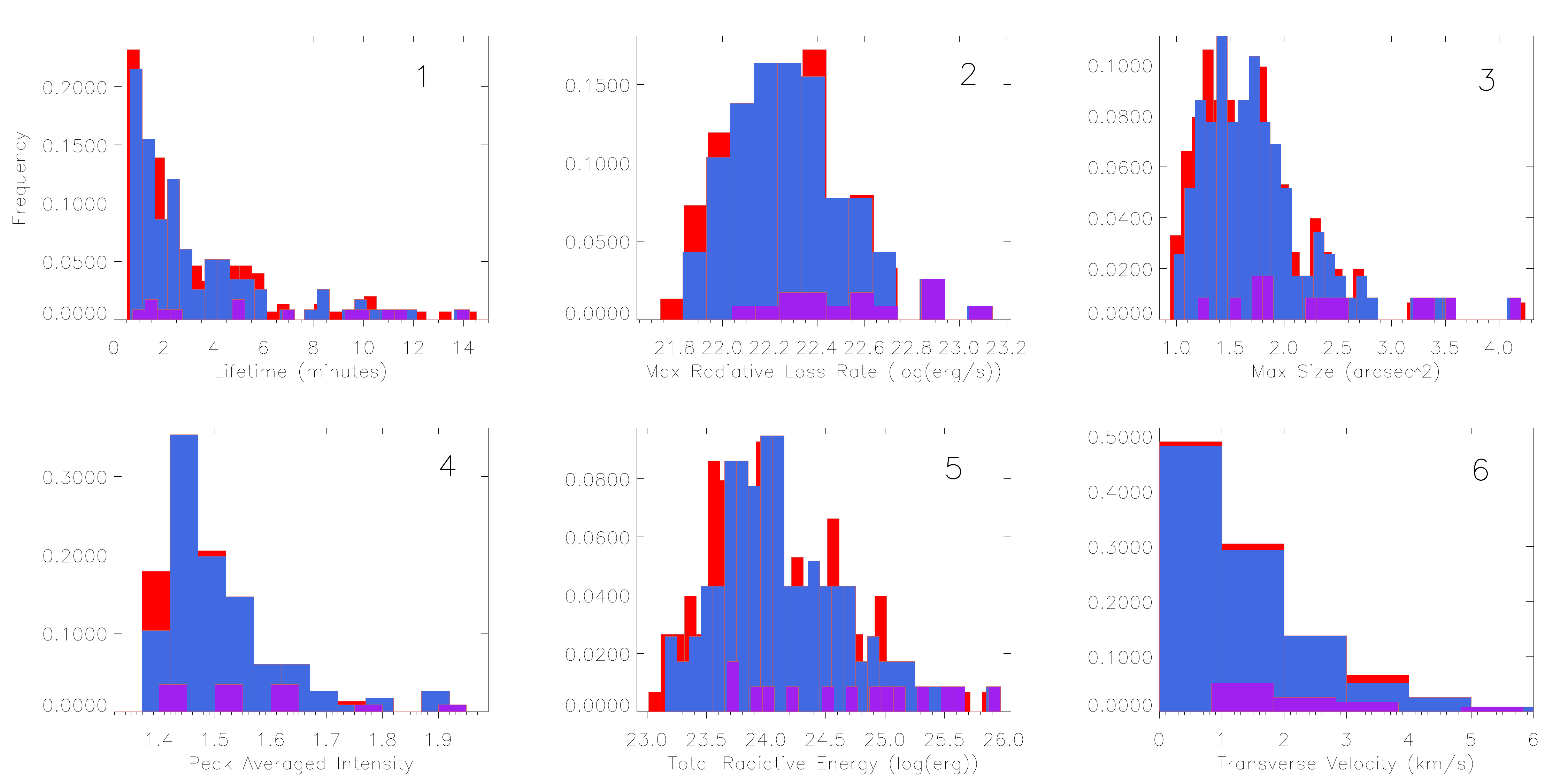}

\caption{Histograms of the EBDATA output with the impulsiveness criterion activated (blue) and deactivated (red) for comparison. The purple bins correspond to the 14 detections from the blue bins which were chosen for inversions (see Section 4).}
\label{Figure2}
\end{figure*}

The output of EBDATA provides the user with sizes, positions, and mean intensity values relative to the averaged background for each detection, along with an EB detection cube showing where each detection is, and its tracking number on the dataset. The outputs can be used to determine lifetimes, apparent transverse velocities, maximum detected intensities (averaged peak intensity over all detected pixels) and sizes (see Fig.~\ref{Figure2}).

Running this detection and tracking algorithm we find 116 EBs in the entire H$\alpha$ dataset. For comparison, the algorithm was also ran with the impulsivity criteria switched off. This resulted in 151 EB detections, though when comparing the statistical characteristics of the 2 sets of results, they were very similar. The additional detections appeared to be short-lived, small-scale, and with lower average intensity (see panels 1, 3 and 4 of Fig.~\ref{Figure2}). This provides some reassurance that the impulsivity criterion removes pseudo-EBs. 

The energy of each detection was also estimated. This was done by adopting the method of \cite{geo}, who adopt the expression for radiative loss rate per unit volume from \cite{nagai} as:
\begin{equation}
\varepsilon_{rad} \simeq a(T)n^{2}\chi g(T)
\end{equation}
where $\chi$ is the ionisation degree, from \cite{brown}, which requires estimates of the local density and temperature. Assuming $n \simeq 10^{12}$ cm$^{-3}$ and that $T \simeq 10^4$ K, the ionization degree for the EB was estimated to be $\chi \simeq 0.89$. Using \cite{nagai} to get the value of a(T) $\simeq 0.1$ for $T \simeq 10^4$ K, they find that the radiative loss rate per unit volume is $\varepsilon_{rad} \simeq 0.89$ ergs cm$^{-3}$ s$^{-1}$. Using this, a peak radiative energy rate could be calculated via: 
\begin{equation}
P_{rad} = \varepsilon_{rad}fV_{EB}
\end{equation}
where $f$ is the filling factor (assumed to be unity) and $V_{EB}$ is the maximum volume of the EB. The volume of the EB was gathered by taking the apparent area encasing the brightening of the EB, and assuming a constant depth of 100 km. Under the assumption that the EB has an equal rising and decaying phase, with a total EB lifetime of D, the total energy could be calculated as:
\begin{equation}
E_{rad} \sim \frac{P_{rad} D}{2}
\end{equation}
The method above is used with the output of EBDATA, using an estimated vertical extent for the EB  of $\approx$ 400~km for equation (2). This is a rough average of the temperature bump widths from semi-empirical modelling \citep{fang, berlicki, li} and observed peak extents of EBs \citep{wat1, Nelson3}. Temperature and density estimates are taken from \cite{geo} which are needed to estimate the net radiative loss rate. The total energy (panel 5 of Fig.~\ref{Figure2}) is then estimated by integrating all calculated energies from the beginning to the end of the detection.  The peak energy rates (panel 2 of Fig.~\ref{Figure2}) are 10$^{22}$ - 10$^{23}$ ergs s$^{-1}$ with total radiative energies ranging between 10$^{23}$ - 10$^{26}$ ergs.

Our algorithm was adapted to allow comparison with previous detection algorithms. \cite{Nelson} found 3570 EBs in a 58\farcss x 58\farcss, 90 minute long dataset of a sunspot. The majority of these detections have since been classified as pseudo-EBs \citep{rutt}. Adapting our algorithm to match their criteria yields 3294 detections, most of which are extremely short lived, small-scale, low intensity brightenings. \cite{Nelson} found 0.684 detections per arcsecond$^{2}$ per second, while our adaptation yielded 0.618 detections per arcsecond$^{2}$ per second. \cite{viss} also created a detection algorithm, identifying 139 potential EBs in 2 datasets of 54\farcss x 53\farcss lasting a total of 106 minutes (0.245 detections per arcsecond$^{2}$ per second). For the dataset presented in this paper, the \cite{viss} criteria yielded 130 potential EBs, corresponding to 0.244 detections per arcsecond$^{2}$ per second. 130 detections is fairly similar to the 116 detected using the criteria adopted in this work. When comparing the statistical characteristics of the detections between the 130 and the 116 presented for this paper, they were extremely similar, with the extra detections appearing small in size (thin long brightenings), with a short lifetime, and velocity.
\begin{table}[!h]
\centering

\begin{tabular}{| c | c | c | c |}
\hline
 \textbf{Algorithm}            & \textbf{Detections} & \textbf{Bipoles} (\%)& \textbf{Strong B Fields} (\%) \\
\hline
Reid & 116 & 68  (58.6)& 102 (87.9) \\ \hline
Vissers            & 130 & 75 (57.7) & 114 (87.6)  \\ \hline
Nelson           & 3294 & 874 (26.5) & 2492 (75.6)   \\ \hline
Reid (NI)    & 151 & 77 (50.1) & 131 (86.7)  \\ \hline
\end{tabular}

\caption{The results of our detection and tracking algorithm (EBDATA), using various detection criteria, with relative magnetic field information. NI stands for Non Impulsive where the impulsivity criterion was disabled.}
\label{table1}
\end{table}

As mentioned earlier, our observations included the Fe I 6302.5~\AA\ line in spectro-polarimetry mode. We checked how many detections overlapped areas of strong photospheric line-of-sight magnetic field, and how many overlapped with areas of opposite polarity magnetic flux. This was carried out with a 20 pixel$^{2}$ (roughly 1.2\farcss x 1.2\farcss ) area around the centre of each detection in the Stokes-$V$ signal where we checked for strong signal in both polarities. A threshold was set to help define what is a strong field, and was considered to be anything with an absolute value greater than the standard deviation of the whole field-of-view at -60~m\AA\ from line centre. This wavelength was chosen to correspond best to the peak of the Stokes-$V$ signal, showing the best contrast for line of sight magnetic fields (similar to \cite{viss} but note the difference in spectral resolution). If a sufficient number of pixels (\textgreater 25\%) was found in the box that met this threshold, then a strong field was assumed to exist. If strong fields were present in both polarities then bipoles were considered to be present. The results in Table~\ref{table1} show that including the impulsivity criterion reduces the number of detections, while strengthening the proportion of detections with associated bipoles.

\section{Photospheric Inversions and Magnetic Flux Cancellation}

The NICOLE inversion code \citep{socasnavarro} was used to determine the evolution of magnetic flux at the EB locations. NICOLE is a parallelized code that  solves multi-level, non-LTE problems following the preconditioning approach described in \cite{socas2}, and that allows for inversions of Stokes profiles, which may contain Zeeman-induced polarization, by using response functions combined with standard fitting techniques \citep{socas3}. The inversions require an initial model to be perturbed, which contains parameters such as a temperature profile, line-of-sight velocity, magnetic field vector, density and microturbulence. Our initial guess model is taken to be the FAL-C atmosphere \citep{font}.  The inversion code attempts to minimize the difference between the observations and the synthetic profiles leading to an inverted model of the observed atmosphere. 

The electron and gas pressures are attained from inserting the temperature stratification
into an equation of state with hydrostatic equilibrium imposed and an upper boundary in
electron pressure. As a result, it may not be possible to obtain flows in the inversion
outputs of EBs, as would be expected in the real case \citep{berlicki}. NICOLE
currently has no alternatives other than hydrostatic equilibrium to get pressures in inversion
mode, and while EBs are impulsive dynamical brightenings, the local magnetic flux output should not be affected. 

Isotropic scattering and complete frequency redistribution are also assumed by NICOLE. While NICOLE uses a plane parallel
atmosphere, radiation comes from and scatters to all directions (I- and I+), with each direction “seeing” a different effective atmosphere. The correct radiation field is important when computing the NLTE populations of the different levels. NICOLE supports up to five angles along a Gaussian quadrature (see e.g. section 5.1.2 of \cite{ruttenbook} for further details of such numerical approximation in this context). We selected 3 angles which is a common compromise between accuracy and speed.  


This atmosphere is perturbed, in a depth dependent fashion by the use of response functions, to converge to a point where the synthetic Stokes output is the most similar to the observed profiles for that pixel. 

Due to NICOLE inversions being computationally intensive, and the possibility that some of the weaker, shorter lived EBDATA detections could be pseudo-EBs, not all 116 detections were inverted. This number was reduced by looking into EB appearances in SDO/AIA channels. 

\cite{viss} tested their algorithm on the SDO 1700~\AA\  AIA channel by using an intensity threshold of 5-$\sigma$ above average instead of the 155\%/140\% thresholding for H$\alpha$. They found a much lower number of EBs than in the H$\alpha$ observations, noting that only the more pronounced EBs were detected in 1700~\AA. Here we have adopted a similar approach. Using co-aligned SDO 1700~\AA\ data, a binary map was created for pixels which were 5$\sigma$ above the average intensity in each frame. In addition, EBs which are only detected near the end of the observations are also discounted, as the purpose of this study is to investigate the change in photospheric magnetic fields over time at EB locations. This narrowed down the 116 detections to just 14, 13 of which contained overlapping opposite polarity photospheric magnetic flux, checked via 6302.5~\AA\ Stokes-$V$, with one detection containing an apparent unipolar region. These were then split into 2 subsets, the primary subset containing all detections which show up with the 1700~\AA\ intensity threshold above 6$\sigma$, and a secondary set of detections which pass the 5$\sigma$ thresholding, but not above 6$\sigma$.

We inverted 100 x 100 pixel$^2$ around each detection. Inversions were done for the EBs relative to their detected start times. Six frames were inverted for each EB, beginning 6 scans prior to the detection, in steps of 6 scans, up to 18 frames after the initial detection. The sixth inverted frame was the final detected frame. Only 6 frames were chosen to show the flux changing over time, while not being extremely intensive. The observations were prepared for NICOLE by normalizing the observed profiles to the $\pm$150~m\AA\ averaged values and interpolating the data points of the spectra to a finer grid. The latter allows enough points for NICOLE to fit the synthetic spectra and the usage of the cubic DELO-Bezier formal solver as described in \cite{jaime2}. To maximize sensitivity to the data, the weights for the interpolated spectral-points not corresponding to an observed wavelength were assigned a negligible non-zero weight.

The NICOLE inversions used 3 nodes in temperature, 1 in line-of-sight velocity, and 1 in vertical line-of-sight magnetic flux density (1 node implies that there will be no fitting of these parameters with height). These numbers were chosen to give simple, effective values of magnetic field and temperature, without introducing increased inversion noise, which would then result in fewer successfully inverted pixels in the EB locations. For the final inversions, no nodes were added for transverse magnetic fields to reduce inversion noise. One test inversion was performed with transverse magnetic field components, and it resulted in little transverse fields at the magnetic inversion line, and so we assume here that the majority of the magnetic flux is vertical. This also allows for a direct comparison with the HMI line-of-sight magnetograms. Stray-light was not added, as this produced incorrect fits, and dramatically increased the inversion noise. This is probably due to the high correlation of the stray-light profile with the data itself, allowing for very good but incorrect fits by perturbing the stray-light component alone. Stray-light is a major component of all observations with similar impact on the contrast of both ground-based \citep{scharmer2} and space based observations \citep{dani}. The observed noise, when fitting stray-light in this work, strongly suggests that spatially-coupled inversions (similar to \citep{vannoort} but with an unknown PSF) would greatly constrain the stray-light fitting and thus lead to greatly improved results. NICOLE currently does not have the ability to perform spatially-coupled inversions.

Adding a microturbulence parameter to the fitting was also attempted. This resulted in very similar results to the inversions without microturbulence, but with better fitting of the Stokes profiles. EBs are impulsive events, dominated by large heating and increased magnetic flux in the upper photosphere. While successfully inverted pixels produced better fits, the microturbulence results do not seem to capture such variability and instead lead to less successfully inverted pixels.



\begin{figure*}[!t]
\includegraphics[trim={1.5cm, 13cm, 0, 1cm}]{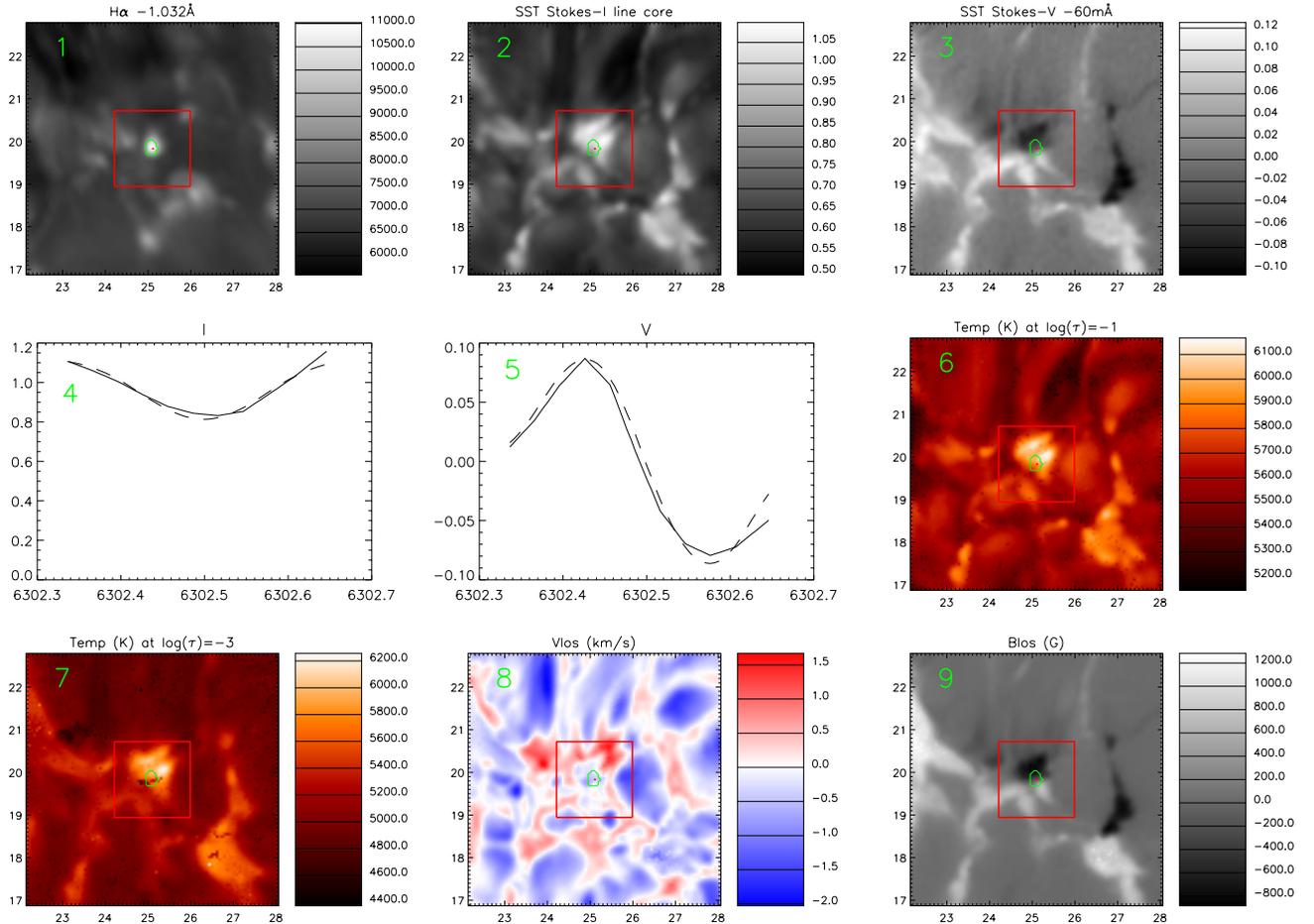}
\caption{Panel 1: The SST observations in H$\alpha$ -1.032~\AA\ of an example EB. Panel 2: Co-spatial Fe I 6302~\AA\ line core Stokes-$I$ imaging. Panel 3: Co-spatial Fe I 6302~\AA\ -60~m\AA\ Stokes-$V$ imaging. Panel 4: The Stokes-$I$ line profiles of the observations (solid) and the synthetic, fitted profiles from the inversions (dashed). The line profiles are taken from a pixel containing strong magnetic field within the red box. Panel 5: The Stokes-$V$ line profiles, co-spatial to Stokes-$I$. Panel 6: The NICOLE output model showing the temperature at  $\log(\tau) = -1$ (corresponding to esitmated height of 220km). Panel 7: The temperature at  $\log(\tau) = -3$ (corresponding to esitmated height of 670km). Panel 8: Line of sight velocity in the upper photosphere (positive = upflow). Panel 9: The line of sight magnetic flux density in the upper photosphere. The green contours show the detected area of the EB from EBDATA.}
\label{Figure_nicole}
\end{figure*}

Fig.~\ref{Figure_nicole} shows example output model data from the inversions with the original SST observations for EBDATA detection 091. At the location of the brightening in H$\alpha$, a temperature rise is visible in the models of roughly 1800K above the FAL-C temperature at the detection footpoint ($\log(\tau) = -1$), (a 490~K increase locally compared to 265 seconds before the detection occurred). The output model also shows a clear bipolar region, shown in the line-of-sight magnetic field map, with absolute flux densities of around 1kG. There is a second bipolar region in the bottom right corner of the observations and output model, which forms detection 092, roughly 13 seconds after this snapshot, and is discussed in the next section. With only 1 node in line-of-sight velocity, we have no information on the velocity gradient as a function of height. There is no evidence for strong velocity fields in the output model. 

All 14 detections were inverted using the procedure described above. The results show flux cancellation at the primary EB locations, with an average rate of (1.01 $\pm$ 0.14) x 10$^{15}$ Mx s$^{-1}$. One of the primary detections shown an increase in magnetic flux. All flux cancelation measurements were made with a 1.8\farcss x 1.8\farcss  box created around the detection location, measured from the initial frame to the final frame of the associated detection. The uncertainties are calculated from the noise of the FOV. The secondary detections show an average flux cancellation rate of (7.73 $\pm$ 1.13) x 10$^{14}$ Mx s$^{-1}$, with two of the secondary detections showing a small increase in local magnetic flux. The increase comes from the negative polarity, while the positive polarity areas experience flux cancellation with every EB detection. The negative polarity magnetic flux increases occur in many of the 14 detections, and is due to areas of negative polarity magnetic flux moving into the area of the EB, continuously feeding the flux into the magnetic inversion line (see attached Movie for example). The positive flux areas are mostly stationary over time. The positive flux cancellation over all 14 detections was at a rate of (1.12 $\pm$ 0.16) x 10$^{15}$ Mx s$^{-1}$, compared to the average total flux cancellation rate of (9.17 $\pm$ 1.26) x 10$^{14}$ Mx s$^{-1}$.

The flux cancellation rate was also measured using co-aligned HMI line-of-sight magnetograms. In HMI, the primary detections had an average flux cancellation rate of (1.25 $\pm$ 0.25) x 10$^{15}$ Mx s$^{-1}$, while the weaker, secondary detections had a lower rate of (6.73 $\pm$ 1.41) x 10$^{14}$ Mx s$^{-1}$. Both HMI measurements and the inversion measurements show similar results, with an average increase in negative polarity flux, and an overall average decrease in positive polarity flux.\\

Magnetic energies were also estimated, from the same area as the flux cancellation measurements, using the following equation: 
\begin{equation}
E_B=(S dl B^2/2\mu)
\end{equation}
where S is the apparent area of the magnetic flux around the detection, and $dl$ is the estimated vertical extent of the EB, assumed to be 400~km (the same as the depth for the radiative energy calculations in Section 3). The magnetic permeability was assumed to be $4\pi$ x $10^{-7}$ N A$^{-2}$. 

 The magnetic energies estimated for the detections were in the region of 10$^{24}$ - 10$^{25}$ ergs. The difference in magnetic energies over time was also calculated, by measuring the magnetic energy difference in the box from the initial detection time to the last inverted frame of the detection. For the primary detections, the magnetic energy difference averaged (3.91 $\pm$ 0.52) x 10$^{24}$ ergs, corresponding to a conversion rate of (2.20 $\pm$  0.29) x 10$^{22}$ ergs s$^{-1}$, while the secondary detections had an average magnetic energy difference of (3.10 $\pm$ 0.46) x 10$^{24}$ ergs with an energy conversion rate of (1.38 $\pm$ 0.28) x 10$^{22}$ ergs s$^{-1}$. \\

Since the magnetic energies were calculated for the 14 stronger EBs, the radiative energies were taken for the same 14 detections for a direct comparison. Directly comparing the magnetic energy differences to the radiative energy of the detections in H$\alpha$ over the same times, we find larger radiative energies to magnetic energy differences on average. This can be explained as some of the EBs are fueled during their lifetimes by new flux emergence or areas of moving magnetic flux. This addition of more magnetic flux will decrease the measured magnetic energy difference within the box, but not the measured radiative energy in H$\alpha$. 

The temperature increase at the detection sites was also investigated. All 14 detections shown enhancements in temperature at the detection site at the frame of initial detection, compared to the same location 6 scans ($\approx$265 seconds) prior to the initial detection. The temperature increases range from 40 - 570~K, with a mean enhancement of 200~K, at the detection footpoints (optical depth of $\log(\tau_{6302}) = -1$, corresponding to an estimated mean physical height of 200~km from the NICOLE output model). The morphology of the temperature enhancements do not appear to follow the shape of the EB in H$\alpha$, and is more characteristic of the EB footpoints.

\section{Ellerman Bomb Pairing}

\cite{zac} noted that EBs can appear and disappear in groups, usually formed in a chain-like pattern, with a separation distance of 1 - 7 \farcss. The EBs had apparent horizontal motions of 0.6 km s$^{-1}$. This pairing of EBs was assumed to be connected to footpoints of magnetic loops caused by emerging flux regions. This fits in with the resistive,  undulatory ``sea serpent'' flux emergence mechanism for EB formation described by \cite{geo, par}. \cite{Nelson3, reid} have also shown that EBs can split apart, forming multiple paired EBs. The splitting occurred with a velocity of roughly 6 km s$^{-1}$, though the separation distance in these cases would be much smaller than that of \cite{zac}. 

\begin{figure*}[!t]
\plotone{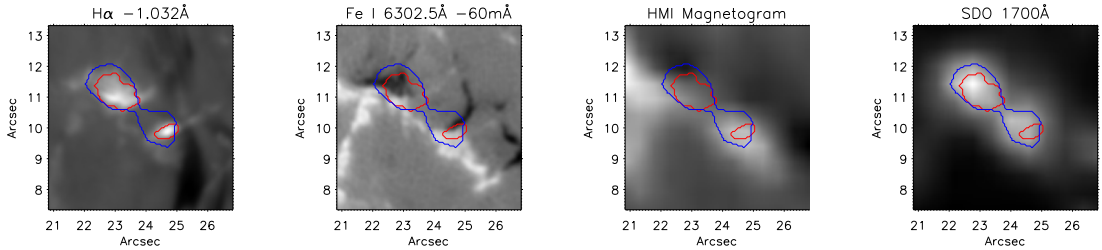}

\caption{Example of an EB pair flaring. The area within the blue contour contains pixels which were greater than 5-$\sigma$ in the 1700~\AA\ continuum, with the red contour highlighting the detections from EBDATA in H$\alpha$.}
\label{Figure3}
\end{figure*}

Our dataset shows 3 instances where EBs are paired, corresponding to roughly 5\% of all EBs in the dataset. Fig.~\ref{Figure3} shows one of those with a mean separation of ~2 \farcss. The origin of the EBs seem to arise from two areas of negative polarity magnetic flux simultaneously emitted from the sunspot, traveling at a measured transverse velocity of 2.5 - 3.5 km s$^{-1}$. The areas of moving magnetic flux then approach a stationary area of opposite polarity magnetic flux, with an EB appearing at both magnetic inversion lines when the opposite polarities meet (see attached Movie). The interaction of the bipolar areas slow the movement of the emitted negative polarity magnetic flux. By comparing the Fe I 6302.5~\AA\ -150~m\AA\ Stokes-$I$ images with the -60~m\AA\ Stokes-$V$/$I$ images of the pairing, it appears as though granular movements affect the movement of the flux regions. Following the appearance of the EBs when the two polarities meet, the faster negative patches for each EB slow down to an apparent velocity of 0.6 km s$^{-1}$ and 0.8 km s$^{-1}$, still moving away from the sunspot. 

\begin{figure*}[!t]
\plotone{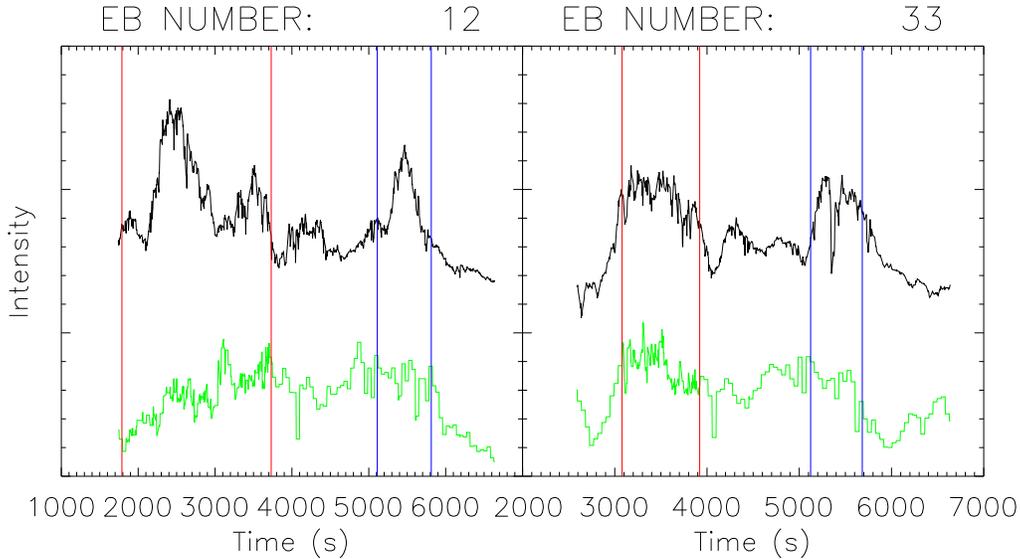}

\caption{Light curves of EB012 (left) and EB033 (right). The black lightcurves (top lines) show the H$\alpha$ -1.032~\AA\ emission over time, inside the boxes described in the text. The green lightcurves (bottom lines) show the Fe I 6302.5~\AA\ -60~m\AA\ Stokes-$V$ absolute intensities over time inside the boxes. The red lines show the start and end detection times of these EBs. The blue lines show the start and end times for the later resurged EB detections 091(left) and 092 (right).}
\label{Figure5}
\end{figure*}

The paired EBs are also detected by EBDATA, labelled as 012 and 033, with lifetimes of 1942 and 841 seconds respectively, and were categorized as primary detections for inversion purposes. Detection 012 has a much longer lifetime due to the area of stationary, positive polarity flux associated with this EB being much closer to the sunspot, and so the bipole forms before detection 033. Fig.~\ref{Figure5} shows the H$\alpha$ -1.032~\AA\ and Fe I 6302.5~\AA\ -60~m\AA\ Stokes-$V$ absolute intensity of each of the detections over time. These lightcurves were measured by placing a 10 pixel$^{2}$ box around the initial detection area for each EB. The EB was then tracked throughout it's lifetime, with the position of the box following the central position of each detection. When the EBs faded and were no longer detected by EBDATA, their associated bipoles were then tracked. The negative polarity flux region within the bipole associated with detection 012 gets fueled by further flux emergence moving out of the sunspot. The fueling of the negative polarity region occurs at least twice where it is noticeable, with roughly 900 seconds between fueling events (the first of which occurs at T=2200s and the negative polarity flux is still emanating out from the sunspot, while the second fueling is noticeable by the large spike in Stokes-$V$ signal at T=3100s). The measured flux cancellation rates were (-2.36 $\pm$ 1.14) x 10$^{14}$ Mx s$^{-1}$ and (1.27 $\pm$ 0.16) x 10$^{14}$ Mx s$^{-1}$ respectively for the detections. The negative value here is due to the fueling of 012. These detections then fade out as the bipolar regions weaken due to the flux cancellation. After 012 and 033 have extinguished, more negative polarity flux emergence from the sunspot fuels the EB areas. 1190 seconds after the last frame detecting 012 and 033, a further 2 detections are also made related to the resurgence of these EBs caused by the flux emergence interacting with the positive polarity existing flux in the area, forming 2 new bipoles. The refueling of the detections is not periodic. 

Flaring in the wings of H$\alpha$ related to the new EBs are labelled as 091 and 092, which have flux cancellation rates of (5.76$\pm$ 0.76) x 10$^{14}$ Mx s$^{-1}$ and (5.90 $\pm$ 0.78) x 10$^{14}$ Mx s$^{-1}$ respectively, measured through inversions. The new detections have lifetimes of 702 and 559 seconds respectively, as shown by the blue bars in Fig.~\ref{Figure5}. The areas of opposite polarity magnetic flux connected to detection 091 lengthen, and by the end of it's lifetime, the magnetic flux tied to the detection contains only negative polarity flux. This lengthening of the areas of opposite polarity flux is similar to that noted previously \citep{reid}, where the bipole connected to the EB was constrained to the intergranular lanes, forming long, thin regions of magnetic flux. Detection 092 disappears at roughly the same time, though the line of sight magnetic fields show a more intricate story. Negative polarity flux connected to 092 seems to split off, with some of the flux staying with the detection, and the rest connecting to a different region of same polarity magnetic flux. This could be due to granular buffeting of the magnetic flux regions, causing a destabilization of the bipoles. The detection loses a large proportion of it's magnetic potential energy, and 092 becomes extinct shortly afterwards. Post extinction, only the positive polarity magnetic fields connected to this detection remain, with the negative polarity magnetic flux region away from detection area still visible.

A second pairing of EBs is also present in the dataset, though in this case the paired system contains a triplet of EBs. Unfortunately these EBs only appear towards the end of the observations, and are not seen in their entirety, and flux cancellation rates cannot be calculated. All 3 EBs are picked up by EBDATA and would have classified as primary EBs, as they passed the SDO 1700~\AA\ test. Each EB within the triplet is formed by negative magnetic flux emerging from the penumbral region of the sunspot and moving out into the surrounding photosphere, where the regions are all met with existing positive polarity areas of photospheric magnetic flux. The apparent transverse motions of the negative polarity patches prior to the EBs flaring have velocities of 2.9 - 3.6 km s$^{-1}$. The EBs in the triplet only begin to appear in H$\alpha$ when the opposite polarity areas meet, 5 minutes before the observations end. It was possible to obtain an estimate for the apparent transverse velocity of the EBs in the triplet. This was found to be 0.6 - 1.1 km s$^{-1}$.

\section{Discussion and Conclusions}

We have developed an Ellerman Bomb Detection Automation and Tracking Algorithm (EBDATA). A skeleton version of this code was adapted to test its functionality in comparison with other existing codes. This resulted in finding a similar amount of EB detections per arcsecond$^{2}$ per second as in previous studies \citep{viss, Nelson}. Using the co-aligned spectro-polarimetric data from the SST, it was shown that the code produced a higher proportion of strong bipolar detections when an impulsive criterion was applied. This reduced the number of false positives, where a moving magnetic region in the photosphere would show up as a brightening in the H$\alpha$ line wings and would fade into the intensity thresholding set in the detection criteria. The algorithm resulted in very similar results to \cite{viss}, with the impulsivity aspect of our algorithm reducing the number of short lived transient brightenings classified as detections. 

Using intensity thresholding of co-aligned SDO 1700~\AA\ data, 14 of the strongest EB detections from EBDATA were selected for inversions. The 6302.5~\AA\ spectro-polarimetric data of these detections were ran through the NICOLE inversion code to find local magnetic flux and temperatures.  The line-of-sight magnetic flux density from the output models of the inversions show that the area around the detections had an average flux cancellation rate of (9.17 $\pm$ 1.26) x 10$^{14}$ Mx s$^{-1}$. Interestingly, when only considering the weaker, secondary set of detections, which had SDO 1700~\AA\ intensities ranging from 5$\sigma$ - 6$\sigma$ above background average, the average flux cancellation rate was (7.73 $\pm$ 1.13)  x 10$^{14}$ Mx s$^{-1}$, indicating that the stronger the intensity in 1700~\AA\ , the stronger the flux cancellation rate. 

Inversion tests show that fits including stray-light vary strongly on a pixel by pixel basis. This shows that spatially-coupled inversions are a highly desirable feature for future development as stray-light is a major component in all observational data.

A comparison of the line-of-sight magnetograms from HMI with our 14 SST detections, shows that the HMI magnetograms at the magnetic inversion line struggle with the low resolution to fully resolve the bipole, and are therefore less reliable for the study of small-scale photospheric magnetic bipoles. The higher spatial resolution SST spectro-polarimetric data provided clearer information on the bipoles with good seeing, which when inverted, gave less noisy, crisper results for flux cancellation rates on the small scale bipoles. However, the HMI measurements seemed to give similar values of flux cancellation to the inverted SST measurements, and could suffice for this purpose. However, without fully resolving the bipole, it would be extremely difficult to ascertain if any fuelling was interfering with the HMI results, or attain any small scale structuring of the bipolar regions under investigation.

EB energies in literature have been reported to be ranging from 10$^{22}$ - 10$^{28}$ ergs. Using equation (2) to work out the radiative energy rates from the 116 detections in H$\alpha$, and integrating these across the detection lifetimes, we find that the resultant energies are 10$^{23}$ - 10$^{26}$ ergs. This is mostly similar to those found by \cite{Nelson}. \cite{geo} found energies of 10$^{26}$ - 10$^{28}$ ergs. These were found to be much higher possibly due to constant lifetime of D=600 seconds, as well as the apparent area of the EBs being much larger than the areas we observe in H$\alpha$. Using the output models from the inversions of the 14 strong EBs, we find magnetic energy differences of 10$^{24}$ - 10$^{25}$ ergs over an average time of 500 seconds,  corresponding to magnetic energy conversion rates of $\approx 10^{22}$ ergs s$^{-1}$. 

Direct comparison of the magnetic and radiative H$\alpha$ energies show that the radiative energy only accumulates to 31.2\% of the magnetic energy difference for the EBs. It is noted here that 5 of the 14 inverted detections were removed from consideration here as they contained apparent refueling in their lifetimes. Without removing EBs which are refueled over their lifetime, the average radiative energy would be higher than the magnetic energy difference. Not all of the magnetic energy which disappears will convert to radiative energy. Some of the magnetic energy will also convert to kinetic energy.

Similar to the flux cancellation rates, the stronger, primary EBs had higher values for magnetic energy conversion rates than the secondary subset. The primary detections had an average peak intensity of 163\% that of the background H$\alpha$, while the secondary detections had an average peak intensity of 148\% (values obtained via the output of EBDATA mentioned in Section 3). This implies that the higher the magnetic energy conversion rate, the brighter the detection appears, and this conversion rate could determine the brightness of the EBs.  

As is evident in Fig.~\ref{Figure5}, flux cancellation and magnetic energy conversion rates may be impacted by the fueling of EBs, and it cannot be claimed that the stronger the measured flux cancellation/magnetic energy, the brighter an individual EB will definitely appear in H$\alpha$/SDO 1700~\AA. Another potential issue is that some EBs have been observed to have their magnetic inversion line lengthen, as seen Section 5, and \cite{reid}. This lengthening may alter the flux cancellation rates, and will most definitely alter the H$\alpha$ wing emission. If the flux cancellation is spread out, the average intensity in H$\alpha$ would also weaken. So while the flux cancellation values could be similar within the box as a whole, if it is more localised inside the measured box, the brightness in H$\alpha$ should also be more concentrated.

The inversions also show a temperature increase at the EB locations. This increase was found to be an average of 200~K at a mean height of 200~km above the photospheric floor ($\log(\tau) = -1$), compared to the local area prior to the detection. This is lower than some previous studies using H$\alpha$ and Ca II 8542~\AA\ data \citep{geo, fang, iso, ach}, which may show an increase in temperature closer to the middle of the EB detection height. The temperature estimates presented in this paper are on the extreme lower end of the newer research provided by co-observations with IRIS using lines more sensitive to higher temperatures \citep{Peter, viss2, kim}, which suggest EBs may have temperature ranges up to 80,000K underneath the chromospheric canopy. The low temperature enhancements we find in comparison are most likely due to our diagnostic line sampling the footpoints of the EBs. The inversions indicate that at a height of 750~km ($\log(\tau) = -3.5$), the mean temperature enhancement rises to 500~K. This result may not be fully reliable due to the low formation height of the Fe I 6302~\AA\ line.  However, this does indicate that the higher areas are heated more than the lower photosphere. This fits with the recent study of \cite{viss2} indicating that the tops of the EBs may be the hottest regions. With only having Fe I 6302.5~\AA\ and H$\alpha$ we are unable to attain any information corresponding to these very high temperatures.

EB pairs have also been found. The pairs were all formed when groups of negative polarity magnetic flux were emitted from the sunspot. The flux travelled away from the sunspot at a velocity of 2.5 - 3.5 km s$^{-1}$. When this flux came into contact with existing, stationary opposite polarity flux, the EBs were formed at the magnetic inversion lines. The movement of the bipoles was much slower than the initial, unhindered negative polarity flux, with a velocity of 0.6 - 1.1 km s$^{-1}$, which is the same as the H$\alpha$ measured transverse velocity of the EBs. This velocity matches well with the previous transverse velocity estimates of \cite{zac}.\\

We would like to thank an anonymous referee for useful comments and suggestions. Armagh Observatory is grant-aided by the N. Ireland Department of Culture, Arts and Leisure. The Swedish 1-m Solar Telescope is operated on the island of La Palma by the Institute for Solar Physics of Stockholm University in the Spanish Observatorio del Roque de los Muchachos of the Instituto de Astrof\'isica de Canarias. The authors wish to acknowledge the DJEI/DES/SFI/HEA Irish Centre for High-End Computing (ICHEC) for the provision of computing facilities and support. We acknowledge support by STFC. This research was supported by the SOLARNET project (www.solarnet-east.eu), funded by the European Commissions FP7 Capacities Program under the Grant Agreement 312495. The research leading to these results has received funding from the European Community's Seventh Framework Programme (FP7/2007-2013) under grant agreement no. 606862 (F-CHROMA). ES is a Government of Ireland Post-doctoral Research Fellow supported by the Irish Research Council.

\end{document}